\documentclass[review]{elsarticle}

\usepackage{lineno,hyperref}

\usepackage{amsmath}
\modulolinenumbers[5]

\journal{Journal of \LaTeX\ Templates}
\usepackage{lineno}
\usepackage{upgreek}
\usepackage{float}
\usepackage{amssymb}

\begin{document}

\begin{frontmatter}

\title{The role of angular slit number and angular slit width on the OAM spectrum}%\tnoteref{mytitlenote}}
% \tnotetext[mytitlenote]{Fully documented templates are available in the elsarticle package on \href{http://www.ctan.org/tex-archive/macros/latex/contrib/elsarticle}{CTAN}.}

% \author{Dina Grace C. Banguilan}%{\fnref{myfootnote}}
% \author{Nathaniel Hermosa}
% \address{National Institute of Physics, College of Science, University of the Philippines, Diliman, Quezon City, 1101 Philippines}

% \author{Dina Grace C. Banguilan* and}
% \author{Nathaniel P. Hermosa}
% \address{National Institute of Physics, University of the Philippines Diliman, 1101 Philippines}
% \address{Corresponding author: banguilandinagrace@gmail.com}

\author[]{Jayson Cabanilla\corref{cor1} }
\ead{jpcabanilla@up.edu.ph}
\author[]{Nathaniel Hermosa}
% \ead{email2@domain.ca}
% \author[add2, add3]{Author3\corref{cor1}}
% \ead{email3@domain.ca}
 
\cortext[cor1]{Corresponding author:}
\address{National Institute of Physics, University of the Philippines Diliman, 1101 Philippines}
% \address[]{\corremail{banguilandinagrace@gmail.com}}
% \address[add2]{Centre for Study of Things}
% \address[add3]{Department of Interests}

% \fntext[myfootnote]{Since 1880.}

% \author[mymainaddress,mysecondaryaddress]{Elsevier Inc}

% \author[mysecondaryaddress]{Global Customer Service\corref{mycorrespondingauthor}}
% \cortext[1]{Corresponding author}
% \ead{banguilandinagrace@gmail.com}

% \address[mymainaddress]{1600 John F Kennedy Boulevard, Philadelphia}
% \address[mysecondaryaddress]{360 Park Avenue South, New York}

\begin{abstract}
The uncertainty principle sets the limit for simultaneous measurements of position and momentum, and its angular analogue is realized through angular diffraction. When a beam is spatially confined by angular slits, the uncertainty in its orbital angular momentum (OAM) increases, leading to the generation of OAM sidebands.
Both the angular slit number and the angular slit width shapes the spatial confinement of the beam. In this study, we investigate this dependence of the OAM  sidebands by obstructing a Gaussian beam with $N$ number of evenly spaced angular slits with angular separation $\Delta\theta_{sep}$. The power distribution among the OAM sidebands exhibits oscillatory behavior as a function of $\Delta\theta_{sep}$. We find that the number of oscillations over the full range $0 \leq \Delta\theta_{sep} \leq 2\pi/N$ is given by the ratio $\frac{|l|}{N}$. Furthermore, each OAM sideband acquires power only when $\frac{|l|}{N}$ takes an integer value, thereby demonstrating the role of the angular slit geometry to the structure of the OAM sidebands.
\end{abstract}


\end{frontmatter}

% \linenumbers

\section{Introduction}

Simultaneous measurement of a particle’s position and momentum is fundamentally constrained by the uncertainty principle \cite{heisenberg2013physical,uffink1985uncertainty}. A similar uncertainty relationship applies to their angular counterparts: angular position and angular momentum . Notably, angular uncertainty also emerges in classical optics, underscoring a fundamental connection between spatial and angular degrees of freedom \cite{franke2004uncertainty,wang2021application}. In classical optics, angular diffraction is the realization of the angular uncertainty, which arises when a light beam passes through angular slits. The angular diffraction generates OAM sidebands, whose distribution is governed by the Fourier relationship between angular position and OAM \cite{yao2006fourier,muino2000introducing,nikolic2011verification}.
The angular position and OAM form a Fourier pair:

\begin{equation}
    A_{l} = \frac{1}{2\pi} \int_{-\pi}^{\pi} \Uppsi(\phi) exp(-il\phi)
\end{equation}

\begin{equation}
    \Uppsi (\phi) = \frac{1}{2\pi} \sum_{-\infty}^{\infty} A_{l} exp(-il\phi)
\end{equation}
where \( \Uppsi(\phi) \) represents the angular amplitude distribution of the field, while \( A_l \) is the complex amplitude of the mode carrying OAM of \( l\hbar \). Here, \( \phi \) is the angular coordinate, and \( l \) is the azimuthal mode index which corresponds to the OAM mode number. In this formulation, angular position and angular momentum are conjugate variables, analogous to linear position and linear momentum.

An angular slit provides a straightforward way to constrain the angular position of a beam. Acting as a perturbation, it modifies the beam’s mode purity by redistributing its OAM content. This confinement in angular position leads to a spread in the OAM spectrum that follows a sinc$^2$-shaped envelope. The broadening reflects a transition from a well-defined OAM state to a coherent superposition of multiple modes \cite{jack2008angular}. The angular slit width influences the extent of this spread: narrower slits produce broader OAM spectrum, while wider slits yield more finite OAM content \cite{torner2005digital, molina2007probing,chen2014quantum}. 

In addition to slit geometry, coherence properties of the incident beam also shapes the spread in the OAM spectrum. Angular partially coherent beams, which exhibit an angular-correlated coherence structure, produce diffraction patterns that vary with their coherence angle. A smaller coherence angle leads to reduced interference and broader OAM sidebands, while a larger coherence angle enhances interference visibility and narrows the OAM spectrum \cite{zeng2025observing}. The use of angular slits to restrict a beam’s position has long been employed for the characterization of its angular momentum properties. For instance, dynamic angular double slits have been used to characterize phase structure through fringe shifts at the angular bisector~\cite{fu2015probing}, and arc-shaped apertures have shown linear variation of the diffraction spot position with the beam’s topological charge~\cite{fu2017detecting}. Sectorial screens have similarly been used to resolve the OAM content of incident beams~\cite{chen2017detecting}.

The angular extent of slits not only defines the spatial structure of the transmitted field but also shapes the OAM spectrum through angular diffraction. While prior studies have explored how evenly-spaced angular slits influence the generation of OAM sidebands, the individual and combined effects of angular slit width and number on the OAM spectrum has received limited attention. In particular, how angular slit width alone redistributes power among OAM sidebands has not been systematically quantified. In this work, we investigate how varying the angular slit width for different numbers of angular slits, influences the generation and distribution of OAM sidebands. We establish a fundamental relationship between the angular slit geometry and the OAM spectrum—paving the way for accurate and adaptable control of structured beams in a wide range of applications \cite{padgett2017orbital,friese1996optical,yang2021optical,gibson2004free,willner2015optical}.

\section{Power expression of the OAM sidebands}
We consider a Gaussian complex field as the mode of the incident beam on the angular slits, while the Laguerre–Gaussian (LG) mode is chosen as the basis for measuring the OAM sidebands of the diffracted field. Owing to its cylindrical symmetry, the LG mode provides a natural representation for structured light fields with well-defined orbital angular momentum. As a paraxial solution to the scalar Helmholtz equation, it is expressed as

\begin{align}
\Uppsi_{pl}(r,\phi,z) &= \frac{C_{pl}^{LG}}{\omega(z)}\left(\frac{r\sqrt{2}}{\omega(z)}\right)^{|l|} \exp\left(-\frac{r^2}{\omega^2(z)}\right) L_p^{|l|}\left(\frac{2r^2}{\omega^2(z)}\right) \nonumber \\
&\quad \times \exp\left(-ik\frac{r^2}{2R(z)}\right) \exp\left(i(l\phi + \varphi(z))\right),
\end{align}
where the normalization constant is given by
\begin{equation}
C_{pl}^{LG} = \sqrt{\frac{2p!}{\pi(p+|l|)!}}.
\end{equation}
We take the lowest radial order $p = 0$, since the radial dependence does not affect the angular diffraction properties \cite{kim1999hermite,jack2008angular}. We are interested in the OAM spectrum immediately after the beam passes through the angular slit, thus, we set $z = 0$. Under these conditions, the LG field simplifies to

\begin{equation}\label{eq:LGfield}
\Uppsi_{0l}(r,\phi) = \frac{C_{0l}^{LG}}{\omega_0} \left( \frac{r\sqrt{2}}{\omega_0} \right)^{|l|} \exp\left( -\frac{r^2}{\omega_0^2} \right) \exp(i l\phi),
\end{equation}
where $\omega_0$ is the beam waist. The fundamental Gaussian beam is simply the LG mode with $p = 0$ and $l = 0$:

\begin{equation}\label{eq:gaussian}
\Uppsi_{00}(r,\phi) = \sqrt{\frac{1}{\pi}} \frac{1}{\omega_0} \exp\left(-\frac{r^2}{\omega_0^2} \right).
\end{equation}
In this formulation, we evaluate the power of each OAM sideband using the overlap integral of the transmitted field and the conjugate of the target OAM mode \cite{hermosa2014nanostep,remulla2019spatial}. The power coupled into a given mode order $l$ is given by
% This approach provides an exact expression for the coupled power, incorporating the appropriate normalization constants and explicitly revealing the dependence on beam parameters such as waist size, slit geometry, and angular alignment. 

\begin{equation}\label{eq:overlap_general}
P = \left| \int_0^\infty \int_0^{2\pi} \Uppsi_{00}^*(r,\phi) M(\phi) \Uppsi_{0l}(r,\phi) \, r \, dr \, d\phi \right|^2,
\end{equation}
where $M(\phi)$ is the mask of the angular slit. 

We use a mask with evenly-spaced angular slits separated angularly by $\Delta\theta_{\text{sep}}$. For a mask composed of $N$ angular slits, the angular separation $\Delta\theta_{\text{sep}}$ spans from $0$ to $\tfrac{2\pi}{N}$. The mask function is defined as  

\begin{equation}
    M(\phi) = 
\begin{cases}
1, & \text{for } (\frac{2\pi m}{N}+\Delta\theta_{sep})\geq\phi \geq (m+1)\frac{2\pi }{N} \\
0,  & \text{otherwise }
\end{cases}
\end{equation}
for $m = 0, 1, \ldots, N-1$. As the angular slit separation is varied, the angular slit width is adjusted accordingly. The angular extent of each slit is given by:

\begin{equation}
\Delta\theta_{\text{slit}} = \tfrac{2\pi}{N} - \Delta\theta_{\text{sep}},
\end{equation}
which also spans from $0$ to $\tfrac{2\pi}{N}$. The overlap equation simplifies to

% \begin{equation}
% P = \left| \int_0^\infty \sum_{m=0}^{N-1} \int_{\frac{2\pi m}{N} + \Delta\theta_{sep}}^{\frac{2\pi (m+1)}{N}} \Uppsi_{00}^*(r,\phi) \Uppsi_{0l}(r,\phi) \, r \, dr \, d\phi \right|^2.
% \end{equation}
% Plugging in equations \eqref{eq:LGfield} and \eqref{eq:gaussian}, the overlap simplifies to

\begin{equation}
P_{0l} = \left| \frac{\sqrt{2}^{|l|}}{\pi \omega_0^{|l|+2}} \int_0^\infty r^{|l|+1} \exp\left(-\frac{2r^2}{\omega_0^2} \right) dr \cdot \sum_{m=0}^{N-1} \int_{\frac{2\pi m}{N} + \Delta\theta_{sep}}^{\frac{2\pi(m+1)}{N}} \exp(i l\phi) \, d\phi \right|^2.
\end{equation}
The integral can be evaluated separately. The radial part yields

\begin{equation}
|I_r|^2 = \left|\int_0^\infty r^{|l|+1} \exp\left(-\frac{2r^2}{\omega_0^2} \right) dr\right|^2 = \left|\left( \frac{\omega_0}{2} \right)^{|l|+2} \Gamma\left( \frac{|l|+2}{2} \right)\right|^2,
\end{equation}
and the angular part gives

\begin{align}
|I_\phi|^2 
&= \left|\sum_{m=0}^{N-1} 
   \int_{\tfrac{2\pi m}{N} + \Delta\theta_{blocked}}^{\tfrac{2\pi (m+1)}{N}} 
   e^{i l \phi} \, d\phi \right|^2 \notag \\[6pt]
&=
\begin{cases}
\displaystyle \frac{2N^2}{l^2} 
  \left( 1 - \cos\!\left( \tfrac{2\pi l}{N} - l\Delta\theta_{sep} \right) \right),
  & l \equiv 0 \pmod{N}, \\[10pt]
0, & \text{otherwise}.
\end{cases}
\end{align}
The result in the angular dependence of the coupled power shows that a non-zero intensity is only measured when the mode index 
$|l|$ is a multiple of the number of angular slits 
$N$, i.e., $|l| \equiv 0 \pmod{N}$.
For a non-integer ratio between $|l|$ and $N$, the integral evaluates to zero, indicating that the modes are suppressed by the angular symmetry. Thus, the final expression for the coupled power simplifies to

\begin{equation}
P_{0l} = \frac{ \left| \Gamma\left( \frac{|l|+2}{2} \right) \right|^2 }{8\pi^2} \frac{N^2}{l^2} \left(1 - \cos\left(\frac{2\pi l}{N} - l\Delta\theta_{sep} \right) \right).
\label{powerfinal}
\end{equation}
The expression above quantifies the power coupled into the \( l \)-th mode after the beam passes through an angular mask composed of \( N \) angular slits, evenly spaced with \( \Delta\theta_{sep} \). The symmetric positioning of the angular slits maximizes interference contrast of each OAM sideband. In this configuration, the transmitted fields travel equal optical path lengths to the observation plane, ensuring that the contributions from each slit remain in phase. By fixing the angular slits in a symmetric distribution, we maximize mode visibility and detection sensitivity for each OAM.

While angular diffraction is known to introduce OAM sidebands, this formulation reveals how the spectral weight of each sideband is shaped by both the \( N \) number of angular slits and the angular width \( \Delta\theta \). In terms of the angular separation, equation \ref{powerfinal} asserts that an OAM sideband only acquires power when the slit separation satisfies the condition
\[
\Delta\theta_{sep} = \frac{2\pi}{N} - \frac{\left(2k + 1\right)\pi}{2l}
\]
where $k \in \mathbb{Z}$. Specific cases can be extracted from the general expression. For instance, when \( N = 1 \), the coupled power is given by
\begin{equation}
P_{0l} = \frac{ \left| \Gamma\left( \frac{|l|+2}{2} \right) \right|^2 }{4\pi l^2} \left(1 - \cos\left(2\pi l- l\Delta\theta_{sep} \right) \right)
\end{equation}
On the other hand, when \( N = l \), the expression simplifies to
\begin{equation} \label{simp}
P_{0l} = \frac{ \left| \Gamma\left( \frac{|l|+2}{2} \right) \right|^2 }{4\pi} \left(1 - \cos\left(2\pi - l\Delta\theta_{sep} \right) \right).
\end{equation}
Both expressions highlight the dependence of the coupled power on the OAM mode index and the angular separation of the slits. Notably, the presence of the cosine term reflects the interference effects arising from the angular arrangement and geometry of the slits. 

The power of the fundamental Gaussian mode is separately obtained. Using the overlap equation with a tailoring Gaussian mode, the power is found to be a function of the angular slit separation $\Delta\theta_{sep}$
\begin{equation}
P_{00} \propto \left| 1 - \frac{N\Delta\theta_{sep}}{2\pi} \right|^2.  
\label{gaussian1}
\end{equation}
The angular separation $\Delta\theta_{sep}$ is bounded between $0$ and $\tfrac{2\pi}{N}$, thus we express $\Delta\theta_{sep}=\tfrac{2\pi}{N}n$, where $0 \leq n \leq 1$. Here, $n=0$ corresponds to a fully open angular slit, while $n=1$ represents the case where the beam is completely blocked. This leads to the power expression
\begin{equation}
P_{00} \propto \left| 1 - n \right|^2.
\end{equation}

\begin{figure}[htbp]
\centering\includegraphics[width=13cm]{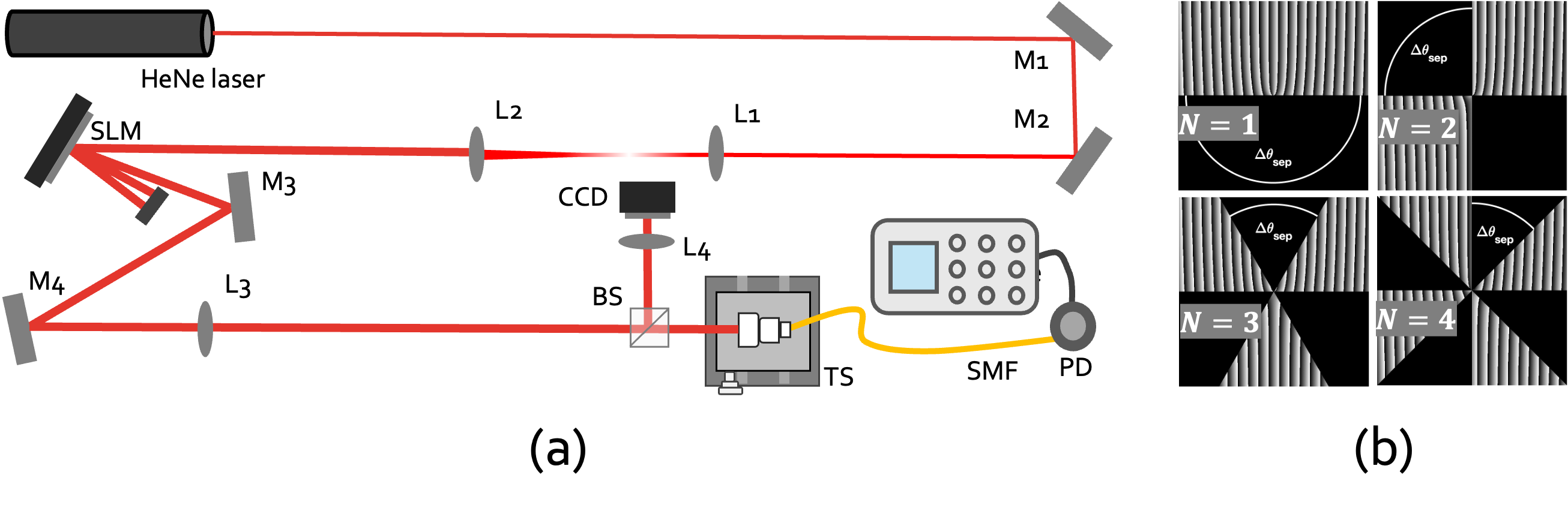}
\caption{(a) Schematic of the experimental setup. A HeNe laser beam is directed onto a spatial light modulator (SLM) displaying a phase hologram of an LG mode superimposed with the angular slits. The diffracted beam passes through a 4f lens system and is subsequently coupled into a single-mode fiber (SMF) mounted on a translation stage (TS). The coupled power is measured using a photodetector (PD). (b) Fork holograms uploaded to the SLM are superimposed with the slits to simultaneously diffract and project the beam onto the target OAM sidebands.}
\label{fig:setup1}
\end{figure}

\section{Methodology}
% \subsection{Diffraction pattern through symmetric angular slits}
We designed an experimental setup to measure the power distributed on the OAM sidebands, following the discussion in the previous section. In Figure~\ref{fig:setup1}, a $632$nm HeNe laser, expanded using collimating lenses $L_1$ and $L_2$, is incident onto a spatial light modulator (SLM) where the diffraction of angular slits and projection on LG modes happen simultaneously. Sample images of the phase holograms are shown in Figure~\ref{fig:setup1}b.
After the SLM, the first-order diffraction was incident on to a 4f lens system. The 4f system ensures that the field of the beam sent onto the single-mode fiber (SMF) corresponds to the beam at \( z = 0 \). Finally, each OAM sideband is measured using the coupled power detected by the photodetector.

\begin{figure}[H]
\centering\includegraphics[width=8cm]{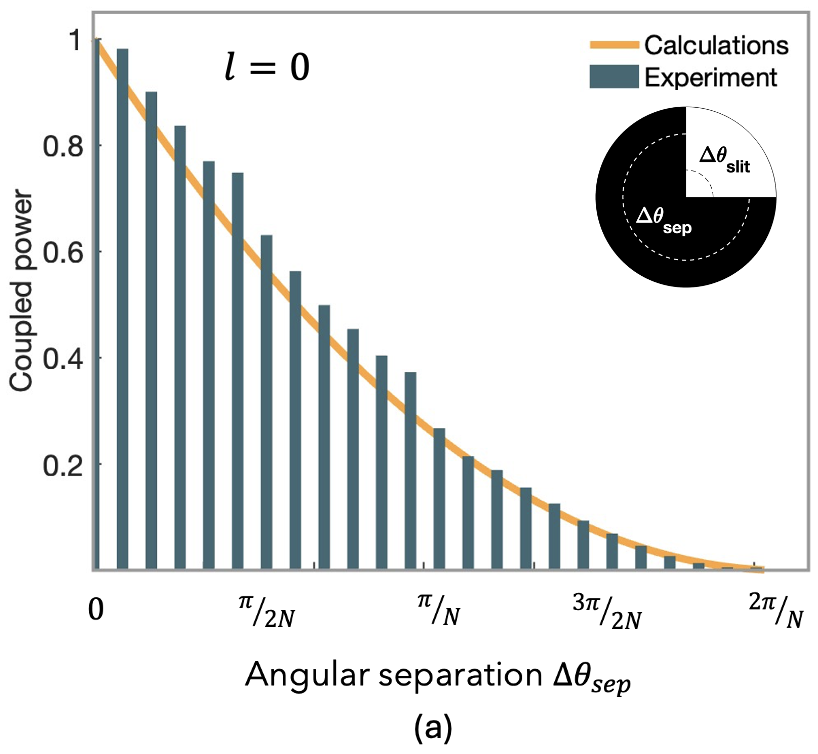}
\caption{Measured (bars) and calculated (solid line) power as a function of angular slit separation \( \Delta\theta_{\text{sep}} \) for the fundamental Gaussian mode, using a single angular slit (\( N = 1 \)). The results show a quadratic decay of the coupled power with increasing angular separation, consistent with the analytical prediction.}
\label{fig:fundamentalmode}
\end{figure}

\section{Results and discussion}
We first measured the power in the fundamental gaussain mode. The angular slit size was varied from \(0\) to \(\tfrac{2\pi}{N}\). Intuitively, the power coupled into the fundamental Gaussian mode reduces with increasing angular separation--or equivalently, decreasing angular slit width. Notably, this reduction exhibits a quadratic decay, as illustrated in Figure~\ref{fig:fundamentalmode}. As shown analytically in Equation~\ref{gaussian1}, the quadratic decay arises from the dependence on the angular separation of the slits.
When the beam passes through an angular slit, the transmitted field is effectively truncated in angle, allowing only a portion of the beam to propagate. The amplitude of the transmitted field in this truncated region scales linearly with the angular slit width since the Gaussian mode has uniform angular amplitude. However, the power coupled into the Gaussian mode is proportional to the square of the amplitude. Consequently, while the field amplitude transmission scales as \(\Delta \theta_{\text{sep}}\), the corresponding coupled power exhibits a quadratic dependence scaled as \(\left(\Delta \theta_{\text{sep}}\right)^2\). The power not coupled into the fundamental mode is partly redistributed into higher-order modes, forming the OAM sidebands of the diffracted beam, while some is absorbed by the opaque portions of the slits.

\begin{figure}[H]
\centering\includegraphics[width=13cm]{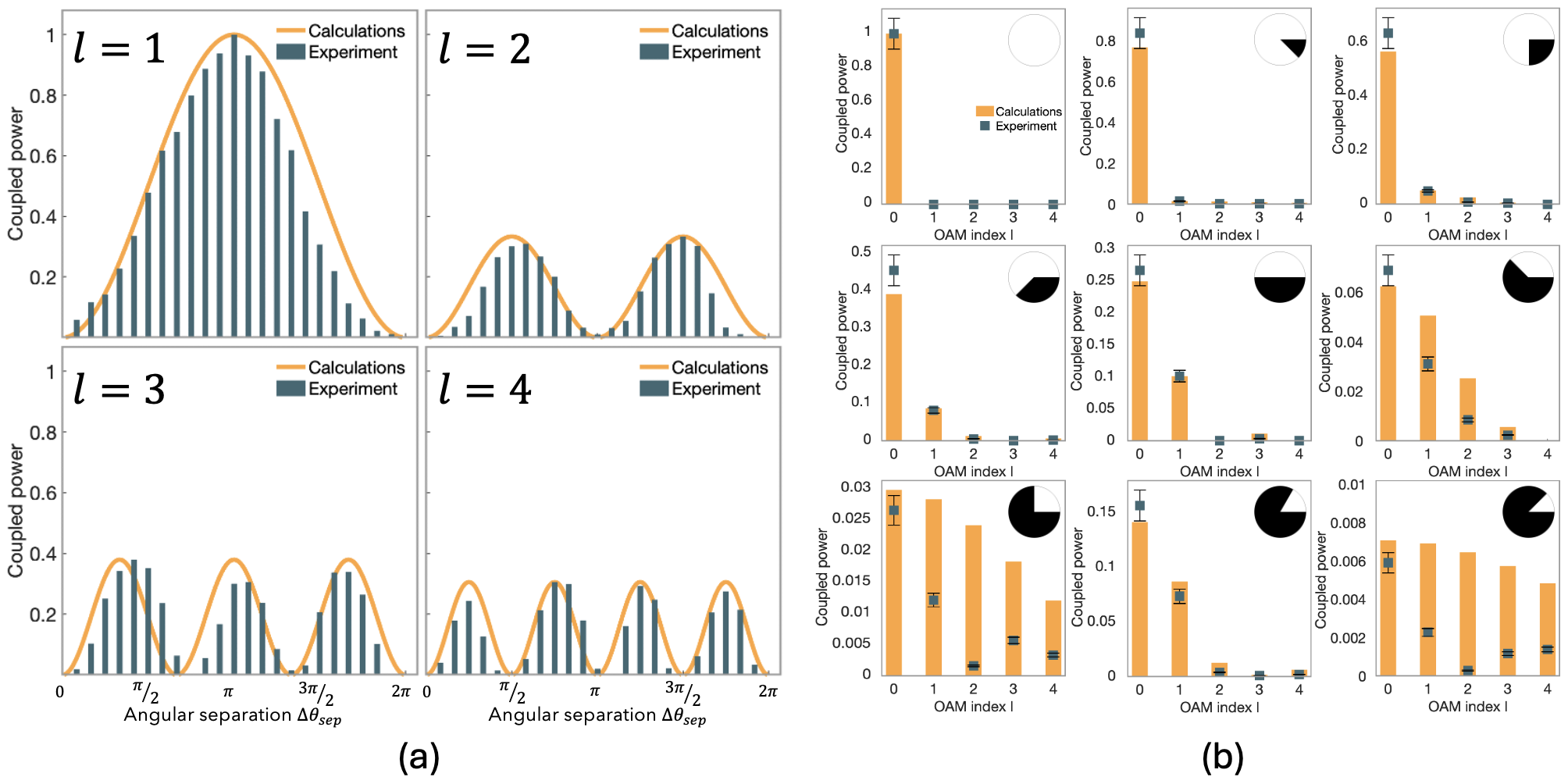}
\caption{(a) Power variation in the OAM sidebands for a single angular slit, \( N = 1 \). Constructive and destructive interferences are observed. The number of power oscillations is equal to the mode of the OAM sideband.
(b) OAM spectrum of a Gaussian beam diffracted by a single ($N=1$) angular slit of varying angular width. As the angular slit width decreases, the power spreads across higher-order OAM modes. This demonstrates how angular confinement of the beam translates into a broadening of the OAM spectrum due to Fourier conjugation between angular position and OAM.}
\label{fig:N_1}
\end{figure}

We measured the power of the OAM sidebands \(l = 1,2,3,4\) of a Gaussian beam diffracted by angular slits. We first considered the case with a fixed slit number of \(N=1\), and the power variation of each OAM sideband is shown in figure~\ref{fig:N_1}a. As the angular slit size increases, each OAM sideband acquires power, although the rate of variation differs. For example, at \(\Delta\theta = \pi\), only odd-order modes contribute to the sideband structure of the diffracted Gaussian beam, while the even-order sidebands are suppressed. The number of local maxima in the power distribution of each sideband—represented by the peaks in the plots—equals the mode index \(l\).  The measured power oscillates with the slit size, and the number of oscillations is exactly equal to the OAM mode of the sideband.

In Figure~\ref{fig:N_1}b, the OAM spectra of the diffracted field produced by a single angular slit with varying widths are shown. For narrower angular slit, the coupled power is more concentrated in the lower-order OAM modes, with the fundamental Gaussian mode (\(l=0\)) being the only mode present when the slit is fully open. As the slit width increases, a broader distribution of modes appears. This behavior is consistent with the uncertainty-principle analog: a wide angular slit transmits a beam closer to the original, unperturbed mode, whereas narrowing the slit causes power to spread into higher-order OAM sidebands due to diffraction effects. As the angular width of the slit decreases, the total coupled power into the fundamental mode diminishes significantly, while the relative power in higher-order modes increases, albeit remaining small. For small angular slit widths, a noticeable breakdown in the experimental results emerges. In this regime, only a minimal fraction of the incident power reaches the detector, resulting in a very weak signal. Under such conditions, experimental limitations such as detector sensitivity has a significant effect. Consequently, the measured data become less reliable and may not follow the power variation predicted by the model. This limitation highlights an interesting avenue for future studies, where more sensitive detection schemes or alternative measurement strategies could enable accurate characterization of the OAM spectrum at extremely narrow slit widths.

\begin{figure}[H]
\centering\includegraphics[width=13cm]{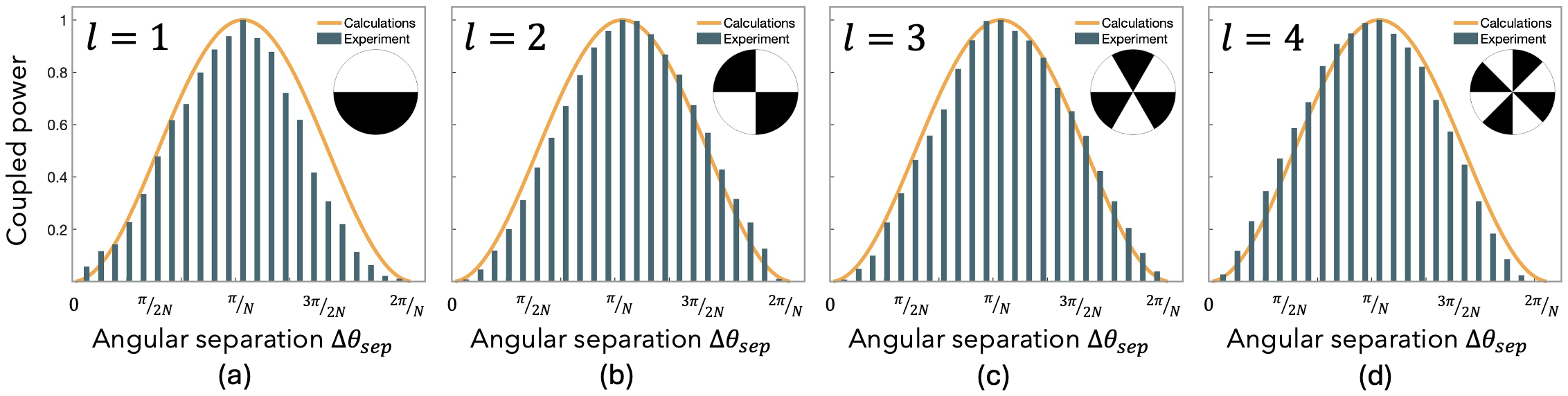}
\caption{ Power variation in the OAM sidebands of a Gaussian beam obstructed by (a)~$N=1$, (b)~$N=2$, (c)~$N=3$, and (d)~$N=4$, with varying angular slit width. Shown here is the case for $N=l$, where only a single maximum, corresponding to one oscillation, is observed in the power variation of each sideband. This behavior arises from constructive interference when the slit periodicity matches the phase winding of the mode.}
\label{fig:N_l}
\end{figure}

We also considered the case for $N=|l|$, with the slit separation constrained to $0 \leq \Delta\theta \leq 2\pi/N$. Although the maximum allowed slit size depends on $N$ (it decreases as $N$ increases), the total open angle across all slits remains the same for every mode, independent of $N$. For example, when \(N=1\), the total open angle is \(\pi\); for \(N=2\), each slit has width \(\pi/2\), giving the same total of \(\pi\) when added.
The power variation are shown in Figure~\ref{fig:N_l}. As the angular separation increases from $0$ to $\frac{2\pi}{N}$, each OAM sideband acquires power. Interestingly, the power on each sideband exhibits a single dominant peak at $\Delta\theta = \frac{\pi}{N}$, regardless of the mode order. This means that there is only a single oscillation in the power when the number of the angular slits matches the mode of the OAM sideband.

\begin{figure}[htbp]
\centering\includegraphics[width=12cm]{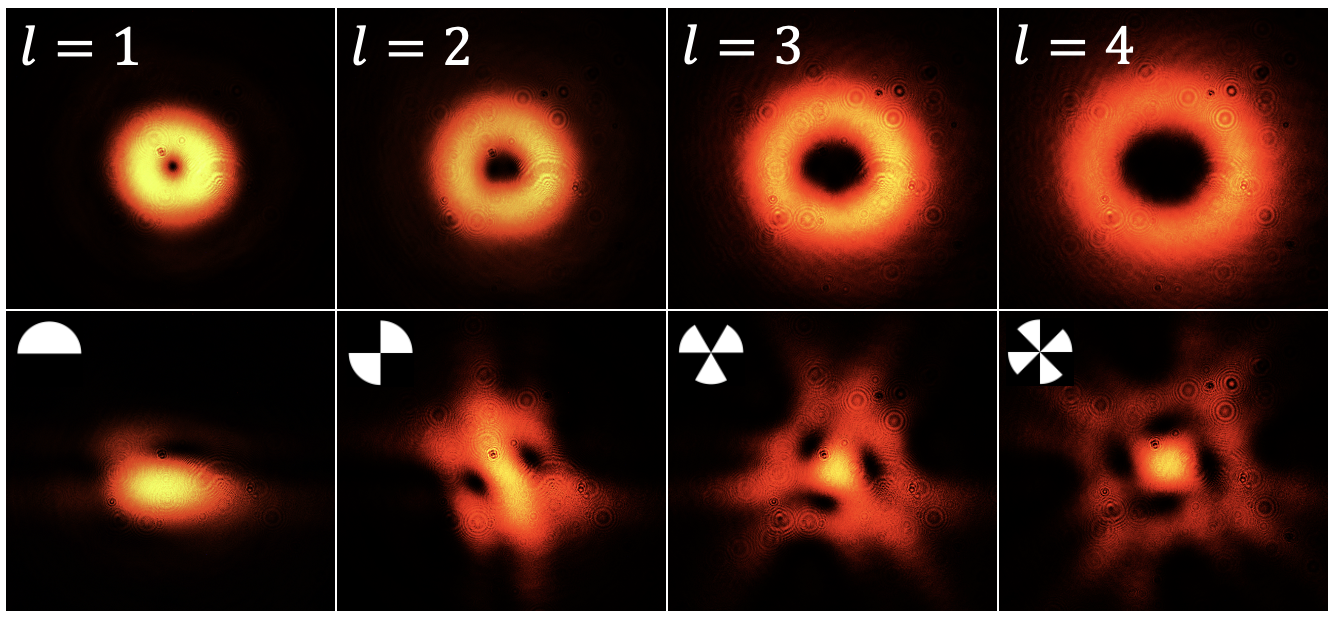}
\caption{Intensity profiles of a Gaussian beam projected onto different LG mode \( l \). The top row displays the projection without obstruction, while the bottom row shows the diffraction patterns immediately after transmission through the angular slits. The number of angular slits is equal to the mode index (\( N = |l| \)) with an angular extent of $ \Delta\theta_{sep}=\frac{\pi}{N}$. A prominent central intensity lobe appears indicating a constructive interference for mode $l$.} 
\label{diffraction}
\end{figure}

Using a charge-coupled device (CCD), sample images of the diffraction patterns were captured for $\Delta\theta_{sep}=\frac{\pi}{N}$ and are shown in Figure~\ref{diffraction}. The first row is when the slit is fully open and mode of the fork hologram is $l$ giving aN LG beam like beam. The second row shows when the slit superposed with the fork hologram has $\Delta\theta_{sep}=\frac{\pi}{N}$. The presence of the central intensity arises from the interference of the beams transmitted through the slits. As seen in the previous cases, when \(N=1\), the number of constructive interference peaks appearing at the center is equal to the ratio \(|l|/N\). Higher-order modes contain more phase windings around the azimuth, which in turn produce additional constructive and destructive interference as the angular slit width varies. For the case \(N=l\), the ratio \(l/N\) equals one, resulting in constructive interference at the center of the diffraction pattern. This produces a pronounced central intensity lobe, which corresponds to the region detected by the single-mode fiber. The detected power is directly proportional to the power contained in the OAM sideband. Thus, when \(N=l\), fully constructive interference occurs at an angular separation of \(\Delta \theta_{\text{sep}} = \pi/N\). This condition arises because the periodicity of the angular mask is synchronized with the mode index \(l\), which defines the azimuthal phase winding. As a result, the phases transmitted through the slits remain coherent and in phase.

\begin{figure}[H]

\centering\includegraphics[width=13cm]{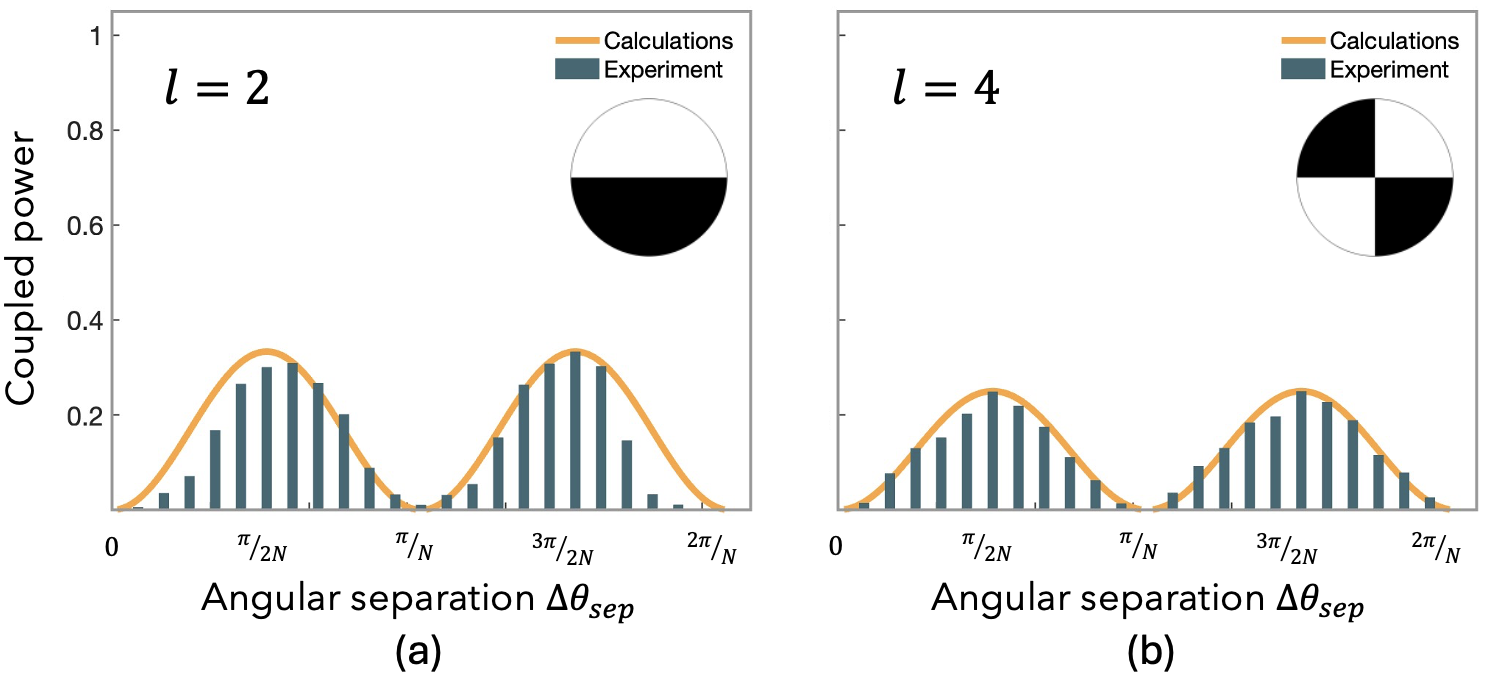}
\caption{Power variation of the OAM sidebands for the case \(N = |l|\), with a Gaussian beam obstructed by (a) \(N = 1\) and (b) \(N = 2\). In both cases, two power oscillations are observed, consistent with the ratio \(|l|/N\).}
\label{fig:N__L__2}
\end{figure}

Taken together, the cases \(N=1\) and \(N=l\) highlight a more general rule. The number of oscillations in the power variation for each OAM sideband is determined by the ratio between the mode order and the number of angular slits, \(\tfrac{|l|}{N}\). As noted in Section~2, only integer values of this ratio yield non-zero power distributions. When \(\tfrac{|l|}{N}\) is an integer, constructive interference reinforces specific sidebands; when the ratio is fractional, the phase contributions from the slits fail to add coherently, resulting in complete destructive interference and zero power in that sideband. We further verified this relationship by measuring the sideband power for \(l=2\) with \(N=1\), and for \(l=4\) with \(N=2\). As shown in Figure \ref{fig:N__L__2}, both conditions produced power variations with two maxima, consistent with the proposed relation. This confirms that a mode sideband acquires power only when \(\tfrac{l}{N}\) is an integer. Otherwise, the transmitted fields from different slits interfere destructively, suppressing power transfer entirely.

 In physical terms, these results show that the angular slit acts as an effective mode filter, with its selectivity determined by the angular slit geometry. A narrower angular slit imposes stronger spatial confinement, which broadens the OAM spectrum and distributes power into multiple OAM sidebands. Conversely, a wider slit preserves more of the incident field, favoring the fundamental Gaussian mode or lower-order OAM states. This filtering behavior arises from the interference of the fields transmitted through the slits, where the angular extent sets the degree of phase coherence among the modes. In essence, the angular slit geometry directly governs which modes are transmitted and how their power are redistributed, thereby providing a simple and tunable mechanism to tailor the modal content of a beam \cite{yang2019manipulation,borghi2016catastrophe,narag2018diffraction}.

\section{Conclusion}
We have shown that the geometry of the angular slits governs the distribution of OAM sidebands. For \(N=1\), the sideband power oscillates with slit size, with the number of oscillations equal to the mode index \(|l|\). When the number of slits matches the mode order, \(N=|l|\), constructive interference produces a single peak at \(\Delta\theta = \pi/N\). More generally, the number of oscillations follows the ratio \(\tfrac{|l|}{N}\), which directly links angular slit number and OAM sideband. Only integer values of this ratio yield measurable sideband power, while non-integer values lead to destructive interference and suppression. These findings show that angular slits can be used as effective mode filters, selectively reinforcing or suppressing OAM sidebands depending on their geometry. By tuning the number and size of the slits, one can determine and control the OAM spectrum of the diffracted field. Such control provides a straightforward means to tailor OAM content for applications in mode analysis, beam shaping, and optical communication. 

\section*{Acknowledgments}
The authors gratefully acknowledge the National Research Council of the Philippines (NRCP) under the Department of Science and Technology (DOST) and the Office of the Chancellor of the University of the Philippines Diliman, through the Office of the Vice Chancellor for Research and Development, for funding support through the Thesis and Dissertation Grant (Project No. 252512 TND).

\end{document}